# Wigner function and intensity moments of spatio-temporal light fields

A. Bekshaev[1]*, O. Angelsky[2,3], C. Zenkova[2,3]

[1]*Physics Research Institute, I.I. Mechnikov National University, Dvorianska 2, 65082, Odesa, Ukraine*
[2]*Research Institute of Zhejiang University-Taizhou, Taizhou City, Zhejiang Province, 318000 China*
[3]*Chernivtsi National University, 2, Kotsyubinsky Str., Chernivtsi 58012, Ukraine*

*\*bekshaev@onu.edu.ua*

**Abstract:** The Wigner distribution function and its spatial-angular moments (intensity-moments) are known as efficient instruments for characterization of complex quasimonochromatic light beams and their transformations. In this paper, the generalization of the WF-based approach to spatio-temporal (ST) light fields (wave packets, short pulses) is considered. It is shown that the ST intensity moments are related with the important characteristics of the wave-packet structure, especially, with the transverse orbital angular momentum (OAM) being a specific feature of the ST optical vortices (STOV). The ST moments' transformations in a paraxial optical system obey simple and unified rules involving the ray-transfer *ABCD*-matrix of the system. On this base, and with simple examples of the OAM-carrying optical pulses, the schemes and mechanisms of the STOV generation and transformation are presented. Examples of non-vortex ST wave packets with the transverse OAM, their possible realizations are discussed as well as the relations between the OAM and the visible pulse rotations. The regular and unified formalism, developed in this paper, can be generalized and applied to more complex situations where the ST field propagates through inhomogeneous and random (scattering) media.

## 1. Introduction

Short and ultrashort light pulses attract a close attention of researchers since the emergence of lasers. Especially important advances were made in recent decades when the ideas of structured light were impressively extended from the spatial to the temporal and spectral domains [1–3]. The most interesting results and concepts concern the fields with essential 4D structuring, where the pulse propagation is coupled with modulations of its spatial and temporal shapes, and the field variations in both domains are inherently interrelated – the spatio-temporal (ST) light fields [3]. Such fields are subjects of the vivid research activity, and the continuous flow of new results essentially enhance our knowledge on the structured light and its properties.

In this context, an important range of issues concerns the laws regulating the ST fields' transformation by various optical systems. Normally, such laws can be easily formulated based on the known approaches widely used in the paraxial optics of conventional monochromatic beams: paraxial wave equation, Kirchhoff – Fresnel integral, geometric-optics rules for the wavefront transformations performed by thin lenses, etc. [4,5,6,7]. In the description of monochromatic paraxial beams and their transformations, very efficient methods were proposed which are based on the properties of optical Wigner functions (WF) [8,9,10,11,12, 13,14] simultaneously representing the field distribution in the spatial and angular (Fourier-conjugated) domains. These methods unite the simplicity and transparency of the ray-optics approach with completeness and consistency of the "true wave" picture of the light propagation, and enable to analyze the transformations performed by arbitrary complex paraxial systems via simple combination of the standard elements characterized by their ray-transfer matrices [15].

The important field spatial characteristics can be described in a standard and unified way via the moments of the WF distribution ("irradiance moments", or "intensity moments" [12,13]). The WF moments offer a generalized but, in many cases, fairly relevant characterization of the beam properties, including their current spatial profile, expected far-field intensity distribution, peculiarities of the internal energy flows, in particular, the orbital angular momentum (OAM) of the beam.

Naturally, one may expect that the mentioned advantages of the WF and intensity-moments instruments can be extended to the ST fields. This possibility looks especially promising in view of the crucial role of the internal energy flows for the properties and evolution of such fields. This role is particularly demonstrative in some novel classes of structured light pulses, for example, the ST optical vortices (STOVs) [16,17,18,19,20,21,22,23,24], which supply an expansion of the optical-vortex concept [7] to the ST domain and are characterized by the phase singularity in the longitudinal plane and by the transverse OAM [25,26,27].

In this paper, we try to realize the above-mentioned possibility, and present a simple version of the WF-based formalism adapted to the ST light fields. The exposition starts with the brief reproduction of the relevant data from the theory of paraxial ST wave packets (Section 2). In Section 3, the ST generalization of the Wigner function is formulated, and its moments of the first and second orders are defined. The peculiar attention is paid to the moments associated with the transverse OAM of the ST fields. The moments' transformations in paraxial optical systems can be described with the help of corresponding generalizations of the ray-transfer matrices discussed in Section 4. As a result, the consistent unified scheme is developed, which is applied for the analysis of the moments' behavior in optical systems composed of arbitrary combinations of the focusing/defocusing elements and homogeneous free-space intervals. In Sections 5 and 6, this scheme is used for examination of the procedures imparting the OAM to ST wave packets of special forms, and for description of fields whose ST profile rotates during propagation ("twisting ST pulses"). In particular, it is shown that the transverse OAM is not always coupled with the STOV structure: like in case of conventional monochromatic optical beams, the non-vortex OAM-carrying fields are possible. Similarly, the non-vortex transverse OAM is not always associated with the intensity-profile rotation. The meaning of the obtained results and potential ways of their further development are discussed in Conclusion.

## 2. Paraxial ST wave packets

Our subject is the light pulse propagating in free space along the axis $z$ (longitudinal); the transverse plane is parameterized by the coordinates $(x, y)$. Nominally, the pulse propagates with the light velocity $c$, and its current shape can be suitably described in terms of the ST variable

$$s = z - ct. \quad (1)$$

Generally, the instantaneous electric field of the pulse obeys the wave equation

$$\nabla^2 \boldsymbol{E} - \frac{1}{c^2}\frac{\partial^2 \boldsymbol{E}}{\partial t^2} = 0$$

(and the same for magnetic field $\boldsymbol{H}$), which constitutes the initial point of the present analysis. In the spirit of ST paraxial approach, it is based on the following assumptions:

(i) the spectral width of the pulse $\Delta\omega$ is small with respect to the central frequency $\omega_0$ which enables to write

$$\boldsymbol{E} = \mathrm{Re}\left[\mathbf{E}(x, y, z, s)\exp(-i\omega_0 t)\right] \quad (2)$$

with $\mathbf{E}(x, y, z, s)$ being a complex function (phasor) slowly varying on time intervals comparable with the oscillation period $T = 2\pi/\omega_0$;

(ii) in turn, the phasor $\mathbf{E}(x, y, z, t)$ can be presented as

$$\mathbf{E}(x, y, z, s) = \mathbf{u}(x, y, z, s)\exp(ik_0 z) \tag{3}$$

where $k_0 = \omega_0/c$ is the central wavenumber, and the complex amplitude $\mathbf{u}$ varies slowly on the distances of the order of wavelength $\lambda = 2\pi/k_0$ (paraxial approximation [4,5,7]). In fact, expressions (2) and (3) represent the zero-order term of a series in degrees of the small parameters

$$(k_0 b)^{-1} \sim (\omega_0 \tau)^{-1} \ll 1 \tag{4}$$

where $b$ and $\tau$ are the characteristic scales of the field variations in space (e.g., beam transverse width) and time (e.g., the pulse duration) [4,5]. For simplicity, we consider the linearly polarized field, $\mathbf{u} = \mathbf{e}_x u$. Then, neglecting the corrections of the order of (4) and higher, the complex amplitude $u(x, y, z, s)$ obeys the usual paraxial wave equation [4]

$$i\frac{\partial u}{\partial z} = -\frac{1}{2k_0}\nabla_\perp^2 u \tag{5}$$

where $\nabla_\perp = \mathbf{e}_x(\partial/\partial x) + \mathbf{e}_y(\partial/\partial y)$, $\mathbf{e}_x$, $\mathbf{e}_y$ and $\mathbf{e}_z$ are the unit vectors of the coordinate axes.

This equation does not contain the variable $s$ (1), which means that, if a certain function $u_a(x, y, z)$ is a solution to Eq. (5), additional solutions can be obtained in the form

$$u(x, y, z, s) = F(s) u_a(x, y, z) \tag{6}$$

where $F(s)$ is an arbitrary function satisfying the natural conditions of physical realizability and compatible with other assumptions (e.g., integrability of $|F(s)|^2$ and the slow variations on the wavelength scale). In particular, $F(s)$ can be taken in the form of a polynomial multiplied by the Gaussian function determining the wave-packet length [27], while the "background" distribution $u_a(x, y, z)$ defines the transverse pulse shape (e.g., Hermite-Gaussian or Laguerre-Gaussian modes [15,28]) slowly evolving with the propagation distance $z$.

In the paraxial approximation [5,7], the complex amplitude $u$ introduced by Eqs. (3), (5), (6), completely determines the optical field and its characteristics. In the current context, the most important characteristics are those associated with the main (longitudinal) energy transfer:

$$p_z = \frac{1}{c} w = \frac{1}{8\pi c}|u|^2. \tag{7}$$

Here, $p_z$ is the $z$-component of the field momentum proportional to the longitudinal energy flow density (Poynting vector component) $c^2 p_z$, $w$ is the energy density, and the whole Eq. (7) describes the field intensity distribution. Additionally, the transverse momentum (transverse energy flow) components are determined by equations

$$p_x = \frac{1}{8\pi\omega_0}\text{Im}\left(u^*\frac{\partial u}{\partial x}\right), \quad p_y = \frac{1}{8\pi\omega_0}\text{Im}\left(u^*\frac{\partial u}{\partial y}\right). \tag{8}$$

In this work, we consider the Gaussian wave packets of the special form where the $y$-dependent part is purely-Gaussian and separable,

$$u(x, y, s) \equiv u(x, s)\exp\left(-\frac{y^2}{2b^2}\right), \tag{9}$$

which permits to omit this $y$-dependent part from the most of calculations (if necessary, its influence on the final results will be accounted for by possible modifications of some normalization constants, and such cases will be specially commented). As to the $(x, s)$-dependent ST part, $u(x, s)$, we make no special assumptions on its form, although, for illustrations, keep in mind the simplest Gaussian STOV [27] with

$$u(x, s) = u_{\text{ST}}^{(1)}(x, s) = \frac{A}{b\sqrt{\pi}}\left(\frac{s}{\zeta} + i\sigma\frac{x}{b}\right)\exp\left(-\frac{x^2}{2b^2} - \frac{s^2}{2\zeta^2}\right) \tag{10}$$

where $\sigma = \pm 1$ denotes the topological charge, $b$ is the transverse size of the field determining the scale of its spatial inhomogeneity (see Eq. (4)), $\zeta$ determines the pulse length, and $A$ is the normalization constant such that the full energy of the STOV (9), (10) equals to

$$W = \int w(x,y,s) dxdyds = \frac{|A|^2}{8\sqrt{\pi}} \zeta. \tag{11}$$

### 3. Wigner function of the ST field and its moments

The method of Wigner function (WF) [8–14] has brought many useful results in paraxial optics. Its idea originates from the fact that the beam field, usually characterized by the transverse complex-amplitude distribution $u(x,y)$, can be equivalently characterized by its Fourier spectrum $U(k_x, k_y)$. In paraxial optics, the latter can be treated as the "angular spectrum" of the field, and the WF "unites" the field distributions in the $(x,y)$ and $(k_x, k_y)$ domains. The WF transformation in a certain optical system can be suitably expressed via the transformation of a geometric ray with the spatial $(x,y)$ and angular $(k_x/k_0, k_y/k_0)$ coordinates. Here, we modify the WF approach in application to the ST fields described above.

First, it should be mentioned that the Fourier spectra of ST fields are well known and widely applicable in the ST optics [3,21,27]. The ST spectrum of the complex amplitude (9) can be expressed by equation

$$U(k_x, k) = \int u(x,s) \exp\left[-i(k_x x + s\Delta k)\right] dxds. \tag{12}$$

Here and further in this paper, the absence of explicit limits of integration means that it is performed over the infinite range of all variables, and the integrals are supposed to converge. Keeping analogy with the paraxial-optics techniques [8–14], the ST WF is determined as

$$I(x,s,k_x,k) = \frac{1}{(2\pi)^2} \int u\left(x + \frac{x'}{2}, s + \frac{s'}{2}\right) u^*\left(x - \frac{x'}{2}, s - \frac{s'}{2}\right) e^{-is'\Delta k - ix'k_x} ds'dx'. \tag{13}$$

It satisfies the condition

$$\int I(x,s,k_x,k) dxdsdk_x dk = \int |u(x,s)|^2 dsdx = \frac{1}{(2\pi)^2} \int |U(k_x,k)|^2 dk_x dk = F \tag{14}$$

where $F$ is the constant related with the field energy. For example, for the model STOV (10),

$$I(x,s,k_x,k) = \frac{b\zeta}{\pi}\left[\frac{s^2}{\zeta^2} + \frac{x^2}{b^2} + 2\sigma\left(\frac{b}{\zeta} sk_x - \frac{\zeta}{b} x\Delta k\right) + b^2 k_x^2 + \zeta^2 \Delta k^2 - 1\right]$$

$$\times \exp\left(-\frac{s^2}{\zeta^2} - \frac{x^2}{b^2} - b^2 k_x^2 - \zeta^2 \Delta k^2\right), \tag{15}$$

and in this case, $F = |A|^2 \frac{\zeta}{b}$. This differs from $\int \left|u_{ST}^{(1)}\right|^2 dxdyds = |A|^2 \sqrt{\pi}\zeta$, following from Eq. (19) of [27], by the multiplier contribution $F_y = \int \exp\left(-\frac{y^2}{b^2}\right) dy = b\sqrt{\pi}$ not included in (15) (see Eq. (9) and the comments beneath it). Generally, $F \cdot F_y = 8\pi W$ where $W$ is the total STOV energy (11).

The most of the WF applications are associated with its moments of the first and second orders [8–14] defined as

$$\mathsf{P} = \frac{1}{F} \int \mathbf{P} I(\mathbf{P}) d^4 \mathbf{P}, \tag{16}$$

$$\mathsf{M} = \begin{pmatrix} \mathsf{M}_{11} & \mathsf{M}_{12} \\ \tilde{\mathsf{M}}_{12} & \mathsf{M}_{22} \end{pmatrix} = \frac{1}{F} \int \mathbf{P}\tilde{\mathbf{P}} I(\mathbf{P}) d^4\mathbf{P}. \tag{17}$$

Here, the ST generalization of the 4D ray column-vector [8–14],

$$\tilde{\mathbf{P}} = \begin{pmatrix} x & s & k_x & \Delta k \end{pmatrix} \tag{18}$$

is introduced, the symbol "~" denotes the matrix or vector transposition (so that $\mathbf{P}$ is the column vector, and the matrix product $\mathbf{P}\tilde{\mathbf{P}}$ in the integral (17) is a 4×4 matrix), and $d^4\mathbf{P} = dsdxdkdk_x$. Remarkably, for calculation of the moments, the explicit expression of the WF (e.g., Eq. (15)) is not necessary; the moments can be found from the "marginal" real-space $u(x,s)$ or spectral $U(k_x, k)$ distributions. In particular, the 4×4 moment matrix $\mathsf{M}$ consists of four 2×2 blocks that can be calculated as [12]

$$\mathsf{M}_{11} = \begin{pmatrix} M_{11} & M_{12} \\ M_{12} & M_{22} \end{pmatrix} = \frac{1}{F} \int \begin{pmatrix} x^2 & sx \\ sx & s^2 \end{pmatrix} |u(x,s)|^2 dsdx; \tag{19}$$

$$\mathsf{M}_{12} = \begin{pmatrix} M_{13} & M_{14} \\ M_{23} & M_{24} \end{pmatrix} = \frac{1}{F} \operatorname{Im} \int u^*(x,s) \begin{pmatrix} x \\ s \end{pmatrix} \begin{pmatrix} \frac{\partial}{\partial x} & \frac{\partial}{\partial s} \end{pmatrix} u(x,s) dsdx; \tag{20}$$

$$\mathsf{M}_{22} = \begin{pmatrix} M_{33} & M_{34} \\ M_{43} & M_{43} \end{pmatrix} = \frac{1}{F} \int \begin{pmatrix} \partial/\partial x \\ \partial/\partial s \end{pmatrix} u^*(x,s) \begin{pmatrix} \frac{\partial}{\partial x} & \frac{\partial}{\partial s} \end{pmatrix} u(x,s) dsdx$$

$$= \frac{1}{(2\pi)^2 F} \int \begin{pmatrix} k_x \\ \Delta k \end{pmatrix} \begin{pmatrix} k_x & \Delta k \end{pmatrix} |U(k_x,k)|^2 dkdk_x. \tag{21}$$

The matrix $\mathsf{M}$ (17) of second moments is symmetric, that is, for its elements $M_{ij}$ belonging to $i$-th row and $j$-th column, the relations take place:

$$M_{ij} = M_{ji}.$$

Like the spatial WF and intensity moments [8–14], elements of the vector $\mathbf{P}$ and matrix $\mathsf{M}$ concisely represent important properties of the field and supply its significant characteristics. Components of the vector $\mathbf{P}$ describe the field "centroid" in the real and Fourier space; elements of the diagonal blocks $\mathsf{M}_{11}$ and $\mathsf{M}_{22}$ generally characterize the field shape in the $(x, s)$ and $(k_x, \Delta k)$ domains, correspondingly [12]. Values of the WF moments depend on the coordinate frame and change when the origin is shifted [12]; to be characteristics of the field 'per se' (independent of the occasional choice of the reference point), the moments should be defined with respect to a certain "intrinsic" center of the field. In the further consideration (and in agreement with the common practice [4,18,19,24,25]) we suppose that this has been done, and the field center corresponds to the point ($x = 0$, $s = 0$). Then, the matrix (17) and its blocks (19) – (21) represent the "central moments" of the considered wave packet.

In this definition, the matrix elements forming the off-diagonal blocks $\mathsf{M}_{12} = \tilde{\mathsf{M}}_{21}$ are related with the internal energy flows [7,29] and are therefore especially interesting in the STOV context. For example, Eqs. (8) and (20 dictate that

$$M_{13} = \frac{8\pi\omega_0}{F} \int x p_x(x,s) dxdyds \tag{22}$$

describes the "weighted" $x$-component of the field momentum (and the corresponding energy flow density $c^2 p_x$). On the other hand, according to Eq. (39) of Ref. [27], the transverse OAM density of the STOV field with respect to the center ($x = 0$, $s = 0$) is $L_y = sp_x - xp_z$, and if the frame is defined so that $\int xp_z dxds = 0$ (the usual condition of the centroid definition, relevant,

for example, for the model STOV (10)), the total OAM of the STOV field can be presented in the form

$$\Lambda_y = \int L_y dxdyds = \frac{1}{8\pi\omega_0} \operatorname{Im} \int su^*(x,s)\frac{\partial}{\partial x} u(x,s) dxdyds$$

$$= \frac{F_y}{8\pi\omega_0} \operatorname{Im} \int u^*(x,s) s \frac{\partial}{\partial x} u(x,s) dxds = \frac{F_y F}{8\pi\omega_0} M_{23} = \frac{W}{\omega_0} M_{23} \quad (23)$$

where Eq. (8) is used, and $M_{23}$ is the element of the moment matrix (17). The relation between the OAM and the off-diagonal elements of the matrix $\mathsf{M}_{12} = \begin{pmatrix} M_{13} & M_{14} \\ M_{23} & M_{24} \end{pmatrix}$ was previously observed in paraxial optics for beams with the conventional longitudinal OAM [7,29], but the STOV situation is peculiar because here the second off-diagonal element $M_{14}$, proportional to $\operatorname{Im} \int u^*(x,s) x \frac{\partial}{\partial s} u(x,s) dsdx$, has no mechanical meaning and gives no OAM contribution.

## 4. Transformation of the ST intensity moments in optical systems

Advantages of the WF formalism in description of paraxial beams are grounded on the simple rules regulating transformations of the spatial and angular ray coordinates [8–14]. Remarkably, the similar rules can be formulated for the ST fields: upon propagation in the first-order optical system, the "ray vector" (18) changes according to the transformation law

$$\mathsf{P} = \mathsf{H}\mathsf{P}^i, \quad \mathsf{H} = \begin{pmatrix} \mathsf{A} & \mathsf{B} \\ \mathsf{C} & \mathsf{D} \end{pmatrix} \quad (24)$$

where the superscript "$i$" denotes the input plane of the system, $\mathsf{H}$ is the 4×4 matrix providing the ST generalization of the transformation matrix [12,15,28,30,31], combined of four 2×2 matrix blocks:

$$\mathsf{A} = \begin{pmatrix} A & 0 \\ 0 & 1 \end{pmatrix}, \quad \mathsf{B} = \begin{pmatrix} k_0^{-1}B & 0 \\ 0 & 0 \end{pmatrix}, \quad \mathsf{C} = \begin{pmatrix} k_0 C & 0 \\ 0 & 0 \end{pmatrix}, \quad \mathsf{D} = \begin{pmatrix} D & 0 \\ 0 & 1 \end{pmatrix}. \quad (25)$$

Here $A$, $B$, $C$, and $D$ are the elements of the usual ray-transfer matrix $\begin{pmatrix} A & B \\ C & D \end{pmatrix}$ of a 2D optical system [15]. For simplicity, we consider the system where all optical elements are situated in the homogeneous space with the refractive index $n = 1$, and the ray-transfer matrix is unimodular (satisfies the condition of unity determinant, $AD - BC = 1$). In this situation, the matrix $\mathsf{H}$ is also unimodular, $\det\mathsf{H} = 1$. The forms of the matrices (25) reflect the difference between the "true" ray coordinate $k_x/k_0$, used traditionally in matrix optics [15,28], and the Fourier-conjugated variable $k_x$ addressed in this work, as well as the fact that the optical system only transforms the "spatial" components $x$ and $k_x$ of the "ray vector" $\mathbf{P}$ (18) but does not affect the "temporal" components $s$ and $\Delta k$.

The main benefits of the moment matrix formalism is that, in combination with the transformation matrix $\mathsf{H}$ (24), it enables the standard and unified description of the moments' transformation in arbitrary first-order optical systems [12,28]:

$$\mathsf{M} = \mathsf{H}\mathsf{M}^i \tilde{\mathsf{H}}, \text{ or } \begin{pmatrix} \mathsf{M}_{11} & \mathsf{M}_{12} \\ \tilde{\mathsf{M}}_{12} & \mathsf{M}_{22} \end{pmatrix} = \begin{pmatrix} \mathsf{A} & \mathsf{B} \\ \mathsf{C} & \mathsf{D} \end{pmatrix} \begin{pmatrix} \mathsf{M}_{11}^i & \mathsf{M}_{11}^i \\ \tilde{\mathsf{M}}_{12}^i & \mathsf{M}_{22}^i \end{pmatrix} \begin{pmatrix} \tilde{\mathsf{A}} & \tilde{\mathsf{C}} \\ \tilde{\mathsf{B}} & \tilde{\mathsf{D}} \end{pmatrix}, \quad (26)$$

or, in more detail,

$$\mathsf{M}_{11} = \begin{pmatrix} M_{11} & M_{12} \\ M_{12} & M_{22} \end{pmatrix} = \begin{pmatrix} A^2 M_{11}^i + 2k_0^{-1} ABM_{13}^i + k_0^{-2} B^2 M_{33}^i & AM_{12}^i + k_0^{-1} BM_{23}^i \\ AM_{12}^i + k_0^{-1} BM_{23}^i & M_{22}^i \end{pmatrix}; \quad (27)$$

$$\mathsf{M}_{12} = \begin{pmatrix} M_{13} & M_{14} \\ M_{23} & M_{24} \end{pmatrix} = \begin{pmatrix} k_0 ACM_{11}^i + (AD+BC)M_{13}^i + k_0^{-1} BDM_{33}^i & AM_{14}^i + k_0^{-1} BM_{34}^i \\ k_0 CM_{12}^i + DM_{23}^i & M_{24}^i \end{pmatrix}; \quad (28)$$

$$\mathsf{M}_{22} = \begin{pmatrix} M_{33} & M_{34} \\ M_{43} & M_{43} \end{pmatrix} = \begin{pmatrix} k_0^2 C^2 M_{11}^i + 2k_0 DCM_{13}^i + D^2 M_{33}^i & k_0 CM_{14}^i + DM_{34}^i \\ k_0 CM_{14}^i + DM_{34}^i & M_{44}^i \end{pmatrix}. \quad (29)$$

These equations supply exhaustive characterization of the ST wave-packet transformation within the accuracy allowed by the WF moments. These enable to trace the modifications of many wave-packet characteristics, embraced by the WF moments, in particular, the OAM (23) associated with the matrix element $M_{23}$.

## 5. OAM-sensitive transformations of the ST fields

The expressions (27) – (29) are valid for arbitrary paraxial ST wave packet. In application to the STOV, the most important consequence is the expression of the matrix element $M_{23}$ (see Eq. (28)) immediately related with the OAM (23):

$$M_{23} = k_0 CM_{12}^i + DM_{23}^i. \quad (30)$$

Equation (30) illustrates some possibilities to manipulate the STOV OAM via transformations performed by combinations of standard optical elements (lenses, mirrors, plane-parallel plates) and free-space intervals, forming an *ABCD*-system [15]. For this purpose, the only important matrix elements are $D$ and $C$, and there are two main sources for the resulting OAM:

(i) If the input wave packet already possesses an OAM ($M_{23}^i \neq 0$), its amount can be adjusted by the proper value of $D$. For example, in a telescopic system ($B = C = 0$), $D = b_i/b$ is the inverse transverse magnification, and

$$M_{23} = \frac{b_i}{b} M_{23}^i, \quad \Lambda_y = \frac{b_i}{b} \Lambda_y^i \quad (31)$$

(cf. Eq. (23)). Accordingly, if the input field is of the simplest form (10), for which $\Lambda_y^i = \frac{\sigma W}{2\omega_0} \frac{\zeta}{b_i}$ (see Eq. (41) of [27]), Eq. (31) gives

$$\Lambda_y = \frac{\sigma W}{2\omega_0} \frac{\zeta}{b}$$

in full agreement with the common theory of the OAM of the STOV [27]. This reasoning shows that the simple telescopic compression (enhancement) of the STOV permits to adjust the value of its OAM in a rather wide range. This effect is quite different from the situation with conventional transverse optical vortices whose symmetric telescopic transformation does not change the longitudinal OAM: the reason is that the usual optical telescope makes no influence on the "temporal" variable $s$ along which the STOV size experiences no compression (enhancement), and for the STOV, the standard telescopic transformation inevitably acts as an asymmetric one.

(ii) Another situation occurs when the input OAM is absent, $M_{23}^i = 0$ in Eq. (30). Then, the OAM can be generated in the process of the pulse focusing (defocusing) by optical systems with $C \neq 0$ provided that the input beam has a non-zero off-diagonal moment $M_{12}^i$.

### 5.1. Spatio-temporal vortex generation

We consider two examples of the OAM generation. The first one is described by Figs. 1 and 2. Here, the input field is a non-vortex wave packet in the form (cf. Eq. (10))

$$u^i(x,s) = \frac{A}{b_i\sqrt{\pi}}\left(\frac{s}{\zeta} + \frac{x}{b_i}\right)\exp\left(-\frac{x^2}{2b_i^2} - \frac{s^2}{2\zeta^2}\right). \quad (32)$$

This distribution is the linear superposition of the (0,0) and (0,1) Hermite-Gaussian modes [28] with the ST coefficients $F(s) \propto s\exp(-s^2/\zeta^2)$ and $\propto \exp(-s^2/\zeta^2)$ and thus represents a physically realizable ST field (see Section 2, Eq. (6)). It is characterized by the oblique nodal line in the $(x, s)$ plane and by the plane wavefront with the edge dislocation (see Fig. 2(a1, a2)).

At the first stage of the OAM-generating transformation, the focusing lens L1 (Fig. 1) of the focal distance $f$ (which corresponds to $C = -1/f$ [15]) adds a parabolic wavefront

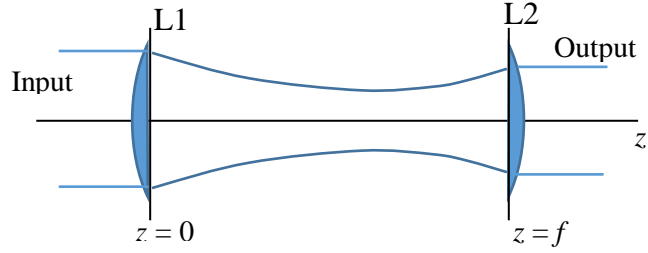

Fig. 1. Scheme of the OAM generation due to the STOV focusing; lenses L1 and L2 are of the same focal distance $f$.

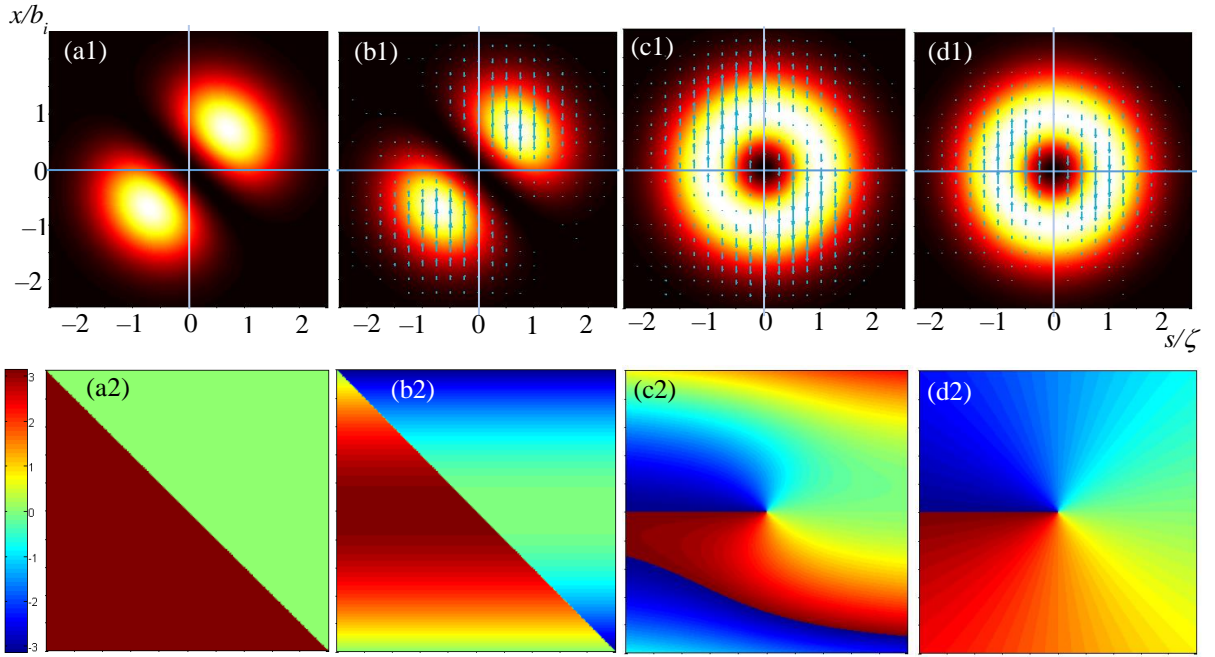

Fig. 2. Evolution of the ST wave packet (32) with $b_i = \zeta = 0.1$ mm, $k_0 = 10^5$ cm$^{-1}$ in the $(x, s)$ plane at different stages of its transformation (see Fig. 1): (a1 – d1) intensity (background color) and transverse energy flows (arrows); (a2 – d2) phase. (a1, a2) Input distribution (32) before the lens L1: the wavefront is flat, transverse energy flows vanish; (b1, b2) distribution (33) just after the lens L1 with $f = 10$ mm ($z = 0$): intensity pattern is the same as in (a1) but the parabolic wavefront associated with the exponential multiplier produces the transverse energy flows towards the $z$-axis; (c1, c2) distribution in the focal plane ($z = f$) of the lens L1: the two-spot input pattern is transformed into the "classic" ring-like STOV (33), but the $x$-momentum distribution is asymmetric due to the divergent wavefront described by the last multiplier of Eq. (35); (d1, d2) the final output distribution after the wavefront correction by the lens L2 (Eq. (35) without the last exponential multiplier).

$$u(x,s,)_{z=0} = u^i(x,s)\exp\left(-\frac{ik_0}{2f}x^2\right). \tag{33}$$

Together with the asymmetric intensity distribution, this wavefront induces disbalance in the *x*-directed momentum density (8) between the right ($s > 0$) and left ($s < 0$) parts of the STOV (Fig. 2(b1, b2)), which, according to Eq. (28) and the first line of Eq. (23), is the source of the transverse OAM [27,26]. Really, Eq. (30) with $M_{23}^i = 0$ gives

$$M_{23} = k_0 C M_{12}^i = -\frac{1}{2}\frac{k_0 b_i}{f}\zeta \tag{34}$$

which means a non-zero transverse OAM (23).

Further propagation behind the lens L1 is described by the Kirchhoff-Fresnel integral [15,32], and at a distance $z = f$, the complex amplitude distribution of the wave packet acquires the form

$$u(x,s)_{z=f} = \frac{k_0}{2\pi i f}\int u(x',s)_{z=0}\exp\left[\frac{ik_0}{2f}(x'-x)^2\right]dx'$$
$$= \frac{kb_i}{if}\frac{A}{\sqrt{\pi}}\left(\frac{s}{\zeta} - i\frac{k_0 b_i}{f}x\right)\exp\left(-\frac{k_0^2 b_i^2}{2f^2}x^2 - \frac{s^2}{2\zeta^2}\right)\exp\left(\frac{ik_0}{2f}x^2\right). \tag{35}$$

This field is illustrated by the images (c1, c2) of Fig. 2: the intensity distribution corresponds to the "classic" ring-like STOV whereas the phase and transverse momentum distributions are distorted by the diverging wavefront associated with the last multiplier of Eq. (35). At the final stage, this distortion is corrected by the lens L2 (Fig. 1) with the same focal distance *f*, after which the "perfect" STOV structure is realized (Fig. 2(d1, d2)).

The whole system of Fig. 1 is characterized by the ray-transfer matrix

$$\begin{pmatrix} A & B \\ C & D \end{pmatrix} = \begin{pmatrix} 0 & f \\ -1/f & 0 \end{pmatrix} \tag{36}$$

which interchanges the spatial and angular (Fourier-conjugated) ray-vector components [15]. Actually, any system with such ray-transfer matrix performs the exact Fourier transform of the initial distribution (32) (one can easily verify this from Eq. (35) after elimination of the last exponential multiplier). The possibility of the OAM generation appears due to the fact that only one pair of the Fourier-conjugated coordinates (*x*, *k*$_x$) of the "ray vector" (18) is involved in the transformation while the second pair (*s*, Δ*k*) remains untouched. This possibility shows essential difference from the conventional vortices with the longitudinal OAM [7], for which the procedures of Fourier transformation, as well as of focusing, cannot generate or eliminate the OAM (although may affect its absolute value if the transformation is asymmetric and implies the different beam compression (enhancement) for different transverse directions [29,33]).

Such schemes of the STOV generation, starting from the non-vortex input wave packet with a nodal line and followed by the Fourier transform, are known and frequently used in practice [19,20]; here we merely illustrate their action based on the moment-matrix framework. The output beam radius of the generated STOV (35) is determined by expression

$$b = \frac{f}{k_0 b_i}. \tag{37}$$

Notably, it can be obtained with the help of the matrix (36) from the generalized transformation rule for the complex beam parameter *q* [15,28]:

$$q = \frac{Aq_i + B}{Cq_i + D} = -f^2\frac{1}{q_i}, \quad \frac{1}{q} = \frac{1}{R} + \frac{i}{k_0 b^2} \tag{38}$$

where $R$ is the wavefront curvature and $b$ is the beam size. In the input plane of the system depicted in Fig. 1, the field (32) possesses a plane wavefront, i.e. $R = \infty$, and $q_i = -ik_0 b_0^2$, whence the first Eq. (38) yields

$$q = -ik_0 b^2 = -if^2 \frac{1}{k_0 b_i^2},$$

which determines the same result as Eq. (37). The OAM of the resulting field (35) can be found directly from the known expression for the STOV OAM (e.g., Eq. (41) of [27]),

$$\Lambda_y = -\frac{W}{2\omega_0}\frac{\zeta}{b}. \tag{39}$$

In view of Eq. (34), (37) and (23), this value exactly coincides with the OAM generated immediately after the lens L1: the OAM is conserved during free propagation and is not changed by the lens L2 because $M_{23} = 0$ for the field (35) before L2.

### 5.2. OAM generation in the non-vortex ST fields

Another example of the transverse OAM generation in the system of Fig. 1 demonstrates that it is not obligatory associated with a STOV. Such situation is well known for monochromatic light beams [7,29,34] where the conventional longitudinal OAM may emerge without phase singularities but merely due to the asymmetric distributions of amplitude and phase; even the productive concept of non-vortex ("asymmetry") OAM was introduced [29]. The analogous non-vortex OAM-carrying optical pulses can be identified in the ST domain. Their existence can be seen from Eq. (30), according to which any asymmetry in the intensity distribution can induce the OAM after the focusing/defocusing procedure. As an example, instead of (32), we consider the wave packet with the initial complex amplitude distribution

$$u^i(x,s) = \frac{A}{b_i\sqrt{\pi}}\left(1 + g\frac{s}{\zeta}\frac{x}{b_i}\right)\exp\left(-\frac{x^2}{2b_i^2} - \frac{s^2}{2\zeta^2}\right) \tag{40}$$

where $g$ is the real constant. This function is a superposition of the fields (6) combining the HG solutions of Eq. (5) with $s$-dependent coefficients and thus, like Eq. (32), also represents a physically realizable ST field; however, any other wave packet with $M_{12}^i \neq 0$ can be taken, for example, tilted Gaussian ST pulses [35,36]. Transformation of the field (40) to the OAM-carrying form can be performed by the same Fourier-transforming optical system of Fig. 1 that was employed in application to the field of Eq. (32). The process of the OAM generation is illustrated by Fig. 3. It looks similar to that of Fig. 2 but in all stages the field singularities are absent or inessential (as the $\pi$-jumps of phase at the peripheral areas of images (b2), (c2)), and the light intensity at the center ($x = 0$, $s = 0$) is not zero but possesses a maximum.

Initially (Fig. 3(a1, a2)), the distribution (40) is characterized by the asymmetric intensity distribution, with the non-zero off-diagonal element of the moment matrix (19),

$$M_{12}^i = \frac{g}{2 + g^2/2}b_i\zeta \neq 0, \tag{41}$$

and the plane wavefront (with some inessential $\pi$-jumps at the periphery); the transverse energy flows are absent. At the first stage, after focusing by the lens L1, the parabolic wavefront along the $x$-direction is introduced (Fig. 3(b1, b2)), and the situation becomes similar to that of the astigmatic Gaussian beams with non-coinciding symmetry axes of the amplitude and phase distributions [7,29]. At this stage, the OAM is imparted to the field according to Eq. (30) ($C = -1/f$):

$$\Lambda_y = \frac{W}{\omega_0}M_{23} = -\frac{W}{\omega_0}\frac{k_0}{f}M_{12}^i = -\frac{W}{\omega_0}\cdot\frac{g}{2 + g^2/2}\cdot\frac{k_0}{f}b_i\zeta. \tag{42}$$

Note that according to Eq. (42), the maximum absolute value of the OAM is reached at $g = \pm 2$, and equals $\Lambda_y = \dfrac{W}{2\omega_0}\dfrac{\zeta}{f}k_0 b_i$ which coincides with the vortex OAM of Eq. (39).

Further transformations do not change the OAM but form the desired intensity and phase structure. After passing the distance $z = f$ (in the focal plane of the lens L1), the wave-packet complex amplitude is described by the relation

$$u(x,s)_{z=f} = \frac{kb_i}{if}\frac{A}{\sqrt{\pi}}\left(1 - i\frac{k_0 b_i}{f}\frac{xs}{\zeta}\right)\exp\left(-\frac{k_0^2 b_i^2}{2f^2}x^2 - \frac{s^2}{2\zeta^2}\right)\exp\left(\frac{ik_0}{2f}x^2\right). \qquad (43)$$

It is characterized by the symmetric intensity distribution but the wavefront is diverging due to the last exponential multiplier of (43) (Fig. 3(c1, c2)). This diverging component is corrected by the lens L2, after which the field with symmetric intensity distribution and smooth regular asymmetric wavefront is formed (Fig. 3(d1, d2)).

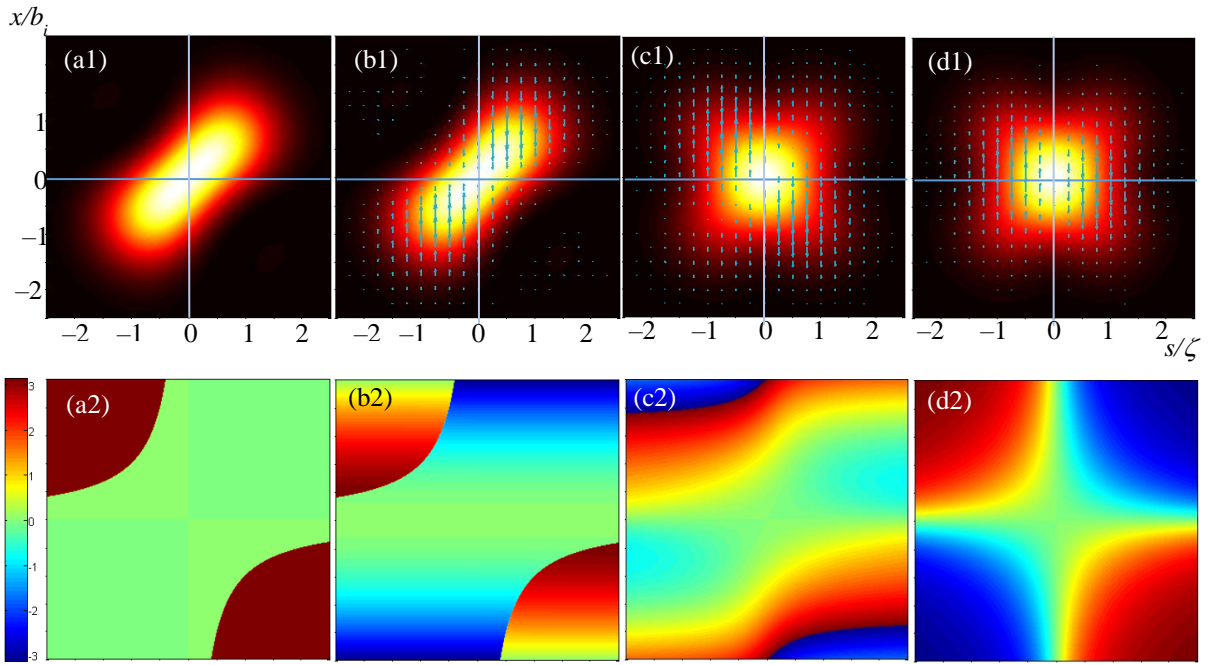

Fig. 3. Evolution of the ST wave packet (40) with $b_i = \zeta = 0.1$ mm, $g = 1$, $k_0 = 10^5$ cm$^{-1}$ in the $(x, s)$ plane at different stages of its transformation (see Fig. 1): (a1 – d1) intensity (background color) and transverse energy flows (arrows); (a2 – d2) phase. (a1, a2) Input distribution (40) before the lens L1: the wavefront is flat, transverse energy flows vanish; (b1, b2) distribution after the lens L1 with $f = 10$ mm ($z = 0$): intensity pattern is the same as in (a1) but the parabolic wavefront formed by the lens produces the transverse energy flows towards the $z$-axis; (c1, c2) distribution in the focal plane ($z = f$) of the lens L1 (cf. Eq. (42)): the asymmetric input pattern is transformed to the symmetric form, but the $x$-momentum distribution is asymmetric due to the divergent wavefront; (d1, d2) the final output distribution after the wavefront corrected by the lens L2.

## 6. Moments-based description of the ST field rotation

Like in the case of paraxial beams [7,12,29], the immediately observable ST field intensity pattern is described by the moment matrix $\mathsf{M}_{11}$ (19). The shape of the ST wave packet in the $(x, s)$ plane can be characterized by the "intensity ellipse" determined by equation

$$(x \quad s)\mathsf{M}_{11}^{-1}\begin{pmatrix}x\\s\end{pmatrix}=\frac{M_{22}x^2+M_{11}s^2-2M_{12}xs}{\det\mathsf{M}_{11}}=1. \tag{44}$$

Eigen values of the matrix $\mathsf{M}_{11}$ determine the ellipse semi-axes, their orientation is specified by the angle

$$\theta=\frac{1}{2}\arctan\frac{2M_{12}}{M_{22}-M_{11}}. \tag{45}$$

Upon free propagation, the transformation of the moment matrix (19) – (21) can be described by Eqi. (27) – (29) for a free-space interval of the length $z$, for which $A = D = 1$, $C = 0$, $B = z$ [12,15]. Accordingly,

$$\begin{pmatrix}M_{11}(z) & M_{12}(z)\\ M_{12}(z) & M_{22}(z)\end{pmatrix}=\begin{pmatrix}M_{11}^i+2k_0^{-1}zM_{13}^i+k_0^{-2}z^2M_{33}^i & M_{12}^i+k_0^{-1}zM_{23}^i\\ M_{12}^i+k_0^{-1}zM_{23}^i & M_{22}^i\end{pmatrix}, \tag{46}$$

$$\begin{pmatrix}M_{13}(z) & M_{14}(z)\\ M_{23}(z) & M_{24}(z)\end{pmatrix}=\begin{pmatrix}M_{13}^i+k_0^{-1}zM_{33}^i & A_{14}^i+k_0^{-1}zM_{34}^i\\ M_{23}^i & M_{24}^i\end{pmatrix},$$

and the matrix $\begin{pmatrix}M_{33}(z) & M_{34}(z)\\ M_{43}(z) & M_{43}(z)\end{pmatrix}$ characterizing the far-field properties [11–13], expectedly keeps invariant ($z$-independent).

In particular, these results show the OAM conservation $M_{23}(z)=M_{23}^i$ and enable to describe the intensity-ellipse rotation during the wave-packet propagation that can be considered as a distinctive attribute of the rotating, or "twisting" [37,38] ST pulses (see Fig. 4). In terms of the intensity moments, this process is characterized by the derivative of the orientation angle (45), which, according to Eqs. (45), (46), equals to

$$\frac{d\theta}{dz}=\frac{1}{k_0}\frac{M_{23}(M_{22}-M_{11})+2M_{13}M_{12}}{(M_{11}-M_{22})^2+4M_{12}^2}. \tag{47}$$

The ST field "twist" means that the angle $\theta$ (45) varies monotonously upon free propagation, i.e. the "rotation velocity" (47) differs from zero and preserves a constant sign within a certain interval of $z$.

In examples considered earlier [37,38], the tilted ST pulses rotate in agreement with their transverse OAM, which leads to supposition that the OAM can be an attribute and a characteristic of the rotation. Equation (47) shows that the relation between the OAM and the rotation is not so direct. Since its denominator is always positive (excluding the degenerate symmetric situations with $M_{11}=M_{22}$, $M_{12}=0$, where the intensity-ellipse rotation makes no sense, and the field profile is also symmetric in terms of the WF second moments), the reasons for the ST pulse twisting can be qualitatively understood from the behavior of the numerator. Hence, one can see that generally there are two sources of twist:

(i) non-zero OAM $M_{23}\ne 0$ (23) together with the condition $M_{11}\ne M_{22}$;

(ii) non-zero "weighted" transverse energy flow $M_{13}$ (22) coupled with the non-zero off-diagonal ST moment $M_{12}$ (19).

In practice, various combinations of the mentioned factors may occur so that any immediate association between the transverse OAM and the ST-pulse rotation becomes elusive. This is illustrated by examples of the focused ST pulses presented below.

### 6.1. The simple polynomial-Gaussian ST field

Figures 4 and 5 demonstrate the propagation of the ST packet which, in the initial plane $z = 0$, differs from the previously considered example (40) by the additional wavefront curvature:

$$u^i(x,s,z=0) = \frac{A}{b_i\sqrt{\pi}}\left(1 + g\frac{s}{\zeta}\frac{x}{b_i}\right)\exp\left(-\frac{x^2}{2b_i^2} - \frac{s^2}{2\zeta^2} - ik_0\frac{x^2}{2f_c}\right). \quad (48)$$

For instance, this field can be obtained if the packet (40) is focused by a lens with the focal distance $f_c$ but it is reasonable to consider expression (48) as a local form of the ST field propagating in the unlimited free-space interval including both negative and positive values of $z$ (see Fig. 5).

Figs. 4 and 5 testify that in case of $f_c = 10$ cm (which coincides with the ST field between lenses L1 and L2 of Fig. 1, see Section 5.2), the field profile, indeed, rotates in the negative (clockwise) direction in agreement with its negative OAM (42). Moreover, this sense of rotation can be traced outside the region considered in Section 5.2, for $z < 0$ and $z > 10$ cm (see Fig. 5, blue curve). While $z < 10$ cm, angle $\theta$ shows the inclination of the intensity-ellipse minor axis (Fig. 4(a1, a2); at $z < 10$ cm the field profile is symmetric (Fig. 4(c1)) and the intensity ellipse degenerates into the circle with indeterminate inclination. With further propagation (Fig. 4(d1)), the minor and major axes of the ellipse are interchanged, and the angle $\theta$ indicates the orientation of the major axis but the clockwise rotation still continues.

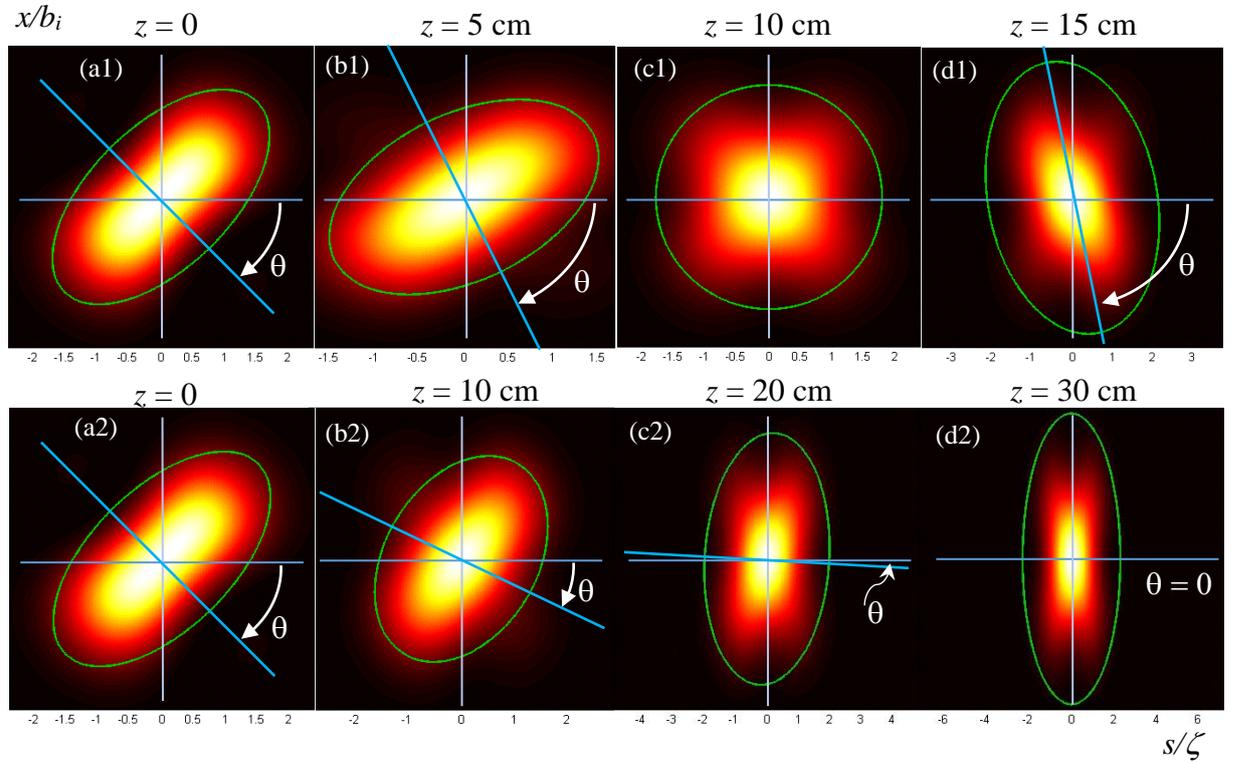

Fig. 4. Simulated intensity profiles of the ST wave packet (48) with $b_i = \zeta = 0.1$ mm, $g = 1$, $k_0 = 10^5$ cm$^{-1}$ in the $(x, s)$ plane at different stages of its free propagation; green contours indicate the shapes and orientations of the intensity ellipses (44), angle $\theta$ specifies the ellipse inclination. Upper row (a1 – d1): $f_c = 10$ cm; (a1) $z = 0$, $\theta = -45°$; (b1) $z = 5$ cm, $\theta = -64.4°$; (c1) $z = 10$ cm, $\theta$ is indeterminate; (d1) $z = 15$ cm, $\theta = -79.6°$. Bottom row (a2 – d2): $f_c = 30$ cm; (a2) $z = 0$, $\theta = -45°$; (b2) $z = 10$ cm, $\theta = -30°$; (c2) $z = 20$ cm, $\theta = -3.5°$; (d2) $z = 30$ cm, $\theta = 0°$.

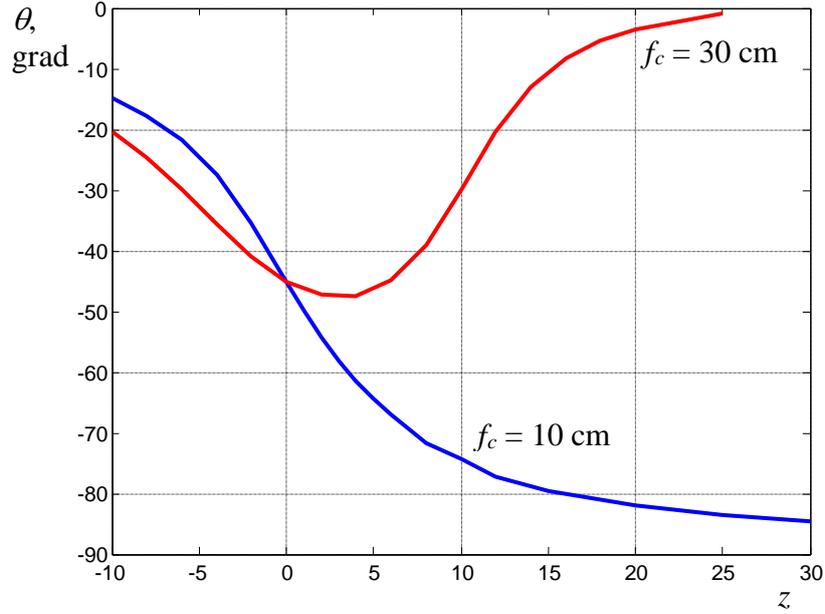

Fig. 5. Dependence of the intensity-ellipse orientation (45) on the propagation distance for the ST wave packet which in plane $z = 0$ has the complex amplitude distribution (48). Numerical data accepted for calculation are the same as in Fig. 4; blue curve: $f_c = 10$ cm (selected ST profiles of the propagating field are illustrated in the upper row of Fig. 4), red curve: $f_c = 30$ cm (selected profiles are illustrated in the lower row of Fig. 4).

Another behavior is observed when $f_c = 30$ cm (Fig. 4(a2 – d2) and Fig. 5, red curve). Here, the "initial" profile at $z = 0$ (Fig. 4(a2)) is the same as for $f_c = 10$ cm (Fig. 4(a1)), and the OAM, also determined by Eq. (42), is again negative. But the negative rotation ($d\theta/dz < 0$) occurs only at $z < 4$ cm, which is illustrated by the red curve in Fig. 5; with further propagation ($z > 4$ cm), the images of Fig. 4(b2 – d2) as well as the ascending segment of the red curve (Fig. 5) explicitly indicate the positive rotation ($d\theta/dz > 0$), opposite to the OAM.

Notably, the similar behavior is known for monochromatic beams whose transverse intensity profile rotates upon the beam propagation but this visible rotation is, generally, not related to the conventional longitudinal OAM of these beams [39]. On the other hand, for asymmetric monochromatic beams, the special concept of the "non-vortex OAM" ("asymmetry OAM") can be introduced that directly characterizes the intensity-ellipse rotation [7,29]. The examples of Section 4 and the discussion presented in this Section show that for the ST fields, the existence of similar criteria, uniquely indicating the intensity-ellipse rotation in the $(x, s)$ plane, looks doubtful (or, otherwise, the expression (47) itself can be considered as such a criterion).

### 6.2. Rotating Gaussian pulse

An additional instructive example is supplied by the "tilted" Gaussian ST packet [35,36]

$$u(x, s, z = 0) \propto \exp\left(-\frac{x^2}{2b_i^2} - \frac{(s + px)^2}{2\zeta^2}\right), \qquad (50)$$

which, after passing a lens, acquires the transverse OAM and shows distinct transverse rotation [37]. Importantly, for the Gaussian wave packet (50), the intensity ellipse (44) characterizes its shape not only "in general" but immediately determines the "genuine" intensity distribution at any cross section of the optical system through which the pulse propagates:

$$\left|u(x,s,z)\right|^2 \propto \exp\left[-\frac{1}{2}(x \quad s)\mathsf{M}_{11}^{-1}(z)\begin{pmatrix}x\\s\end{pmatrix}\right] \tag{51}$$

(here, $z$ denotes the current cross-section of the system). Due to this fact, the moments-based formalism is especially suitable in this case, and the whole ST field evolution can be studied algebraically in terms of the intensity moments, without explicit involvement of the wave equation or Kirchhoff-Fresnel integral, which was inevitable with other approaches [37].

For the "initial" wave packet (50) (prior to focusing), the moment-matrix elements can be directly calculated from the definitions (17), (19) – (21)

$$\mathsf{M}_{11}^i = \begin{pmatrix} M_{11}^i & M_{12}^i \\ M_{12}^i & M_{22}^i \end{pmatrix} = \frac{1}{2}b_i^2\begin{pmatrix} 1 & -p \\ -p & \zeta^2 b_i^{-2} + p^2 \end{pmatrix}; \quad \mathsf{M}_{12}^i = \begin{pmatrix} M_{13}^i & M_{14}^i \\ M_{23}^i & M_{24}^i \end{pmatrix} = 0,$$

$$\mathsf{M}_{22}^i = \begin{pmatrix} M_{33}^i & M_{34}^i \\ M_{43}^i & M_{44}^i \end{pmatrix} = \frac{1}{2\zeta^2}\begin{pmatrix} \zeta^2 b_i^{-2} + p^2 & p \\ p & 1 \end{pmatrix}. \tag{52}$$

Focusing of the packet (50) means [37] that it enters an optical system including the lens of the focal distance $f_c$, plus the free-space interval $z$, for which $A = 1 - z/f_c$, $B = z$, $C = -1/f_c$, $D = 1$ [15]. By using Eqs. (52) and (27) – (29), all the WF moments can be easily found analytically for arbitrary $f_c$ and $z$. Here we only reproduce those ones that, according to Eqs. (45) and (47), explicitly affect the transformations of the observable intensity profile (51):

$$M_{11}(z) = M_{11}^i\left(1-\frac{z}{f_c}\right)^2 + \frac{z^2}{k_0^2}M_{33}^i, \quad M_{12}(z) = M_{12}^i\left(1-\frac{z}{f_c}\right), \quad M_{22}(z) = M_{22}^i,$$

$$M_{13}(z) = -\frac{k_0}{f_c}\left(1-\frac{z}{f_c}\right)M_{11}^i + \frac{z}{k_0}M_{33}^i, \quad M_{23}(z) = -\frac{k_0}{f_c}M_{12}^i, \quad M_{33}(z) = M_{33}^i - \frac{k_0^2}{f_c^2}M_{11}^i. \tag{53}$$

Eqs. (53) and (51), (52) give an immediate possibility to calculate the profile of the focused tilted pulse with the initial form (50) in any cross section behind the lens. The results shown in Fig. 6 and 7 explicitly confirm the rotational nature of the focused pulse [37] and specify the variations of its orientation $\theta$ (45) during propagation.

These results, obtained by using the WF moments, are in good agreement with the behavior discussed previously [37] (note, however, that, for convenience, the numerical parameters of the packet (50), accepted in Figs. 6, 7, differ from those of [37], and the coordinate $s$ (1) is used instead of the time $t$)). In contrast to the case of Figs. 4(a2 – d2) and 5 (red curve), here the rotation velocity (47), determined by the 1st Eq. (52) and Eqs. (53), preserves its sign and agrees with the transverse OAM (positive, $M_{23}(z) = (k_0/2f_c)pb_i^2 > 0$). This feature is akin to the situation of Figs. 4(a1 – d1) and 5 (blue curve); the only difference is that there the sign of the OAM and the rotation handedness are opposite. Remarkably, in conditions of Fig. 5 (blue curve), absolute value of the "full" rotation angle upon propagation along the whole longitudinal axis does not reach 90° whereas in case of Fig. 7, the "full" rotation angle upon propagation from the lens output ($z = 0$) to the infinity is close to 180°. This is similar to the "integral" rotation of the transverse profile of monochromatic beams associated with the Gouy-phase increment [40].

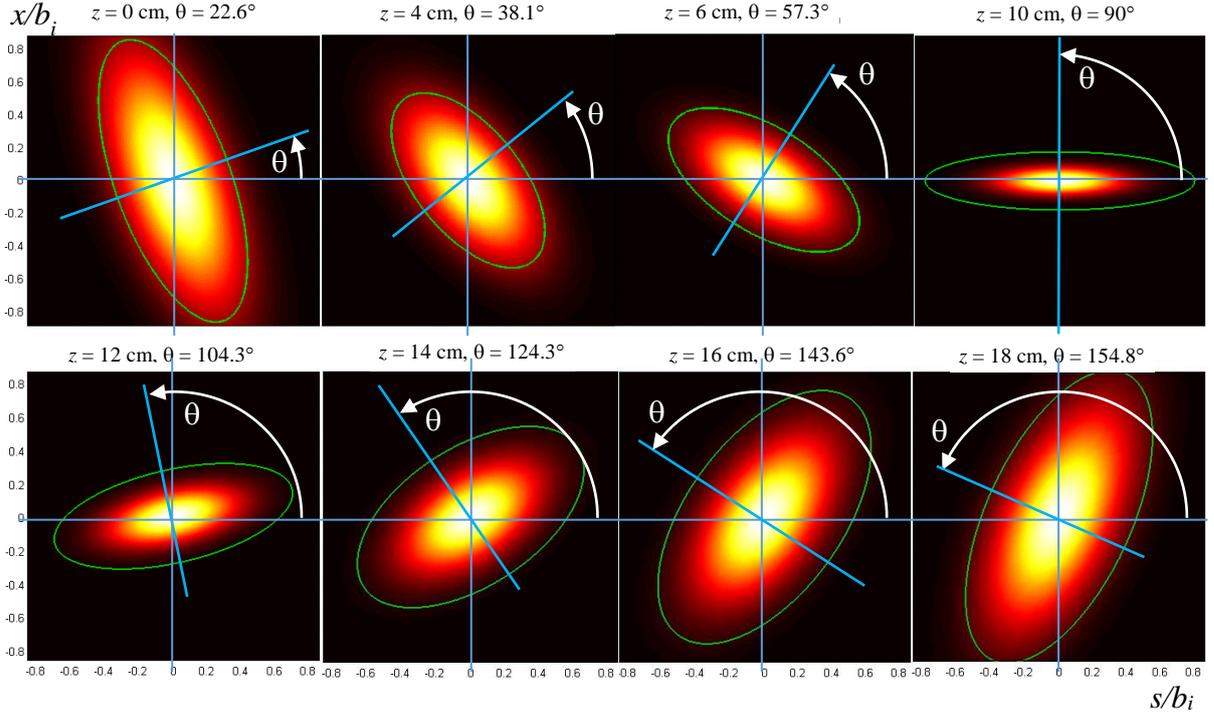

Fig. 6. The intensity profile evolution of the tilted ST packet (50) with $b_i = 0.35$ mm, $\zeta = 0.15$ mm, $p = 0.3$ and the central wavenumber $k_0 = 0.833 \times 10^5$ cm$^{-1}$, after passing a focusing lens with $f_c = 10$ cm, with growing distance $z$ behind the lens. Green contours schematically show the intensity-ellipse orientations, blue lines indicate positions of their minor semi-axes, $\theta$ is the angle of rotation determined by Eq. (45).

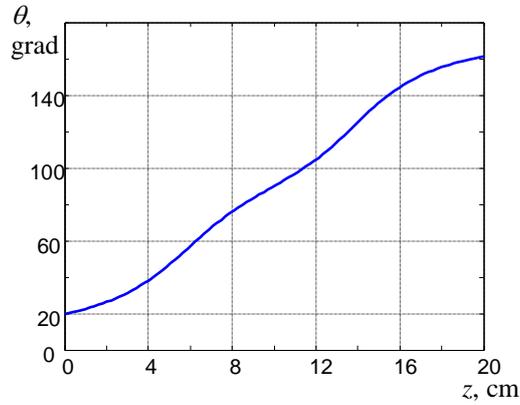

Fig. 7. The intensity-profile orientation $\theta$ (45) of the propagating Gaussian ST packet (50) after focusing by a lens with $f_c = 10$ cm, as a function of the propagation distance $z$. The values of all parameters are the same as in Fig. 6.

## 7. Conclusion

We hope that the materials presented in previous sections convincingly demonstrate the essence and facilities of the WF and moment-matrix formalism in application to the ST light fields. The main conclusion is that these instruments, developed earlier for quasi-monochromatic light beams [8–14], after the minor and physically transparent modifications enable the regular and unified description of the ST-field transformations performed by the common paraxial optical systems. Like their prototypes in the monochromatic-beam theory, the ST intensity moments

(19) – (21) supply a succinct but, in many situations, appropriate characterization of the ST field "shape" and distribution of the internal energy flows. The simple rules for the ST moments' transformations (26) – (29) emerge as a proper adaptation of the equations valid for the elements of the 4×4 spatial moment matrix in astigmatic paraxial systems [12]. These rules employ the well-known ray-transfer *ABCD*-matrices [12,13,15,30,31], introducing their ST generalizations (25).

The ST intensity moments characterize important features and attributes of the ST-field internal structure. One of these attributes is the transverse OAM (23), which is a definitive property of the STOVs, intensively studied during the recent years [16–27]. Due to this circumstance, the results obtained in this paper offer new possibilities for the description and theoretical investigation of various types of the STOV fields, disclosing the nature and mechanisms of their generation and transformation. In this context, the results of Figs. 2, 3, 4 and 6, illustrating evolution of the internal structure of the OAM-carrying fields, indicate the conditions at which the special distributions of the field energy and energy flows can be realized and employed, e.g., in problems of optical manipulation and/or information processing [21].

As a side result of the present research, the theoretical scheme for generation of the non-vortex wave packets with OAM is proposed and analyzed (Section 5.2). It is similar to the recently described procedure of imparting transverse OAM by focusing spatiotemporally coupled ultrashort light pulses [37]. The non-vortex ST OAM-carrying fields are the analogs of the non-vortex monochromatic beams with OAM [29,34] and share some their properties. In particular, in some cases their OAM can be associated with the visible beam-profile rotation during the free propagation (and this rotation of the ST wave packets in the ($x$, $s$)-plane really occurs [37,38]). In Section 6, simple examples of such "rotating", or "twisting" ST fields are considered, and it is shown that the visible rotation of the ST intensity profile is stipulated by a number of factors. The transverse OAM is only one of them, and its "rotational" influence can be enhanced, suppressed and even inverted by specific conditions and parameters of the ST field structure. The WF moments supply a suitable formalism for analysis of such situations, based on the intensity ellipse and its parameters (44) – (47).

Another important feature of the non-vortex OAM is that it is not conditioned by the field topology and can thus be adjusted deliberately in a wide range. On the other hand, the non-vortex ST fields are not topologically stable and are therefore less appropriate for the optical communications and data encoding. However, they still are able to transfer the angular momentum, and in this capacity, the non-vortex OAM-carrying ST fields offer some additional benefits associated with the high energy concentration at the pulse center (in contrast to the STOVs where the energy is accumulated in the off-center ring). This property can be favorable in problems of optical manipulations [21].

Important limitations of the WF-moment methods stem from the requirement of convergence for the definitive integrals (19) – (21); moreover, even if the integrals converge, their values can be very sensitive to the behavior of the field distributions (9), (13) at the far ($x$, $s$)-infinity, where the "physical" field practically vanishes [12]. This hampers the moments' applications to some popular field models, e.g., Bessel fields or edge-diffracted waves [41,42]. In such cases, many characteristics based on the moments' ratios (for example, Eqs. (45), (47)), can still give finite and meaningful results [42] but, generally, additional precautions may be necessary for properly "cutting" the model-field amplitude at the regions where the field intensity is practically imperceptible [43,44]. For Gaussian-based wave packets, considered in this paper, the ST intensity moments are well defined and usable without restrictions.

As a final remark, it should be mentioned that, whereas the WF and moments' definitions (Eqs. (13), (16) – (21)) and their transformations (Eqs. (24) – (29)) are presented in general forms, their applications are considered rather briefly with the special simplified examples (Sections 5, 6). Of course, the potential of the WF-based methods is much wider (see, for example, Refs. [9,10,12,13,14]), and many techniques employing the spatial moments in the monochromatic-beam theory can be applied to ST fields. The most evident way of extension concerns partially coherent and inhomogeneously polarized optical fields, for which the WF and its vector (matrix) generalizations are known as natural and efficient descriptive tools [9,10,12, 13,14]. On the other hand, the rules of the moments' transformation can be modified for more complex optical systems, including inhomogeneous lens-like media [12], random scattering media [45], etc., providing thus additional instruments for analysis of the ST fields in the corresponding conditions. In principle, ways for the further development of the techniques, proposed in this paper, are clear but their realization requires a separate work.